\documentclass[prb,twocolumn,showpacs,amsmath,amssymb,superscriptaddress]{revtex4-2}
\usepackage{amsmath,amssymb}
\usepackage[pdftex]{hyperref,graphicx}
\hypersetup{colorlinks = true, urlcolor = blue, linkcolor = blue, citecolor = blue}
\usepackage{physics}
\usepackage{xcolor}
\usepackage{bm}

\newcommand{\comment}[1]{{}}

\newcommand{\cmmnt}[1]{}

\begin{document}
\title[]{Anisptropic plasmons in threefold Hopf semimetals}
\author{Seongjin \surname{Ahn}}
\email{sahn@cbnu.ac.kr}
\affiliation{Department of Physics, Chungbuk National University, Cheongju 28644, Korea}

\begin{abstract}
Threefold Hopf semimetals are a novel type of topological semimetals that possess an internal anisotropy characterized by a dipolar structure of the Berry curvature and an isotropic energy band structure consisting of a Dirac cone and a flat band. In this study, we theoretically investigate the impact of internal anisotropy on plasmons in threefold Hopf semimetals using random-phase approximation.  In contrast to the classical intuition that isotropy of the energy band dispersion leads to isotropic plasmons in the classical regime (i.e., in the wavelength limit), we find that plasmons in threefold Hopf semimetals exhibit notable anisotropy even in the long-wavelength limit. We derive an explicit analytical form of the long-wavelength plasmon frequency, and numerically demonstrate the validity of our results in a wide range of situations. Our work reveals that the anisotropy of long-wavelength plasmons can reach 25\%, making it experimentally observable. 

\end{abstract}

\maketitle

\section{INTRODUCTION}
Topological semimetals have garnered significant interest in the physics community owing to the unique properties arising from their topological electronic structures \cite{fangMultiWeylTopologicalSemimetals2012,fangTopologicalNodalLine2016,burkovTopologicalSemimetals2016,armitageWeylDiracSemimetals2018a}. A representative example is the Dirac semimetals, which are characterized by a linear band crossing between the conduction and valence bands \cite{armitageWeylDiracSemimetals2018a}. This band crossing serves as a source or sink of the Berry curvature, analogous to a magnetic monopole in momentum space \cite{berryQuantalPhaseFactors1984, fangAnomalousHallEffect2003, unzelmannMomentumspaceSignaturesBerry2021}. These unconventional electronic structures often result in peculiar phenomena, such as the unusual density dependence of long-wavelength plasmon frequencies\cite{hwangDielectricFunctionScreening2007, lvDIELECTRICFUNCTIONFRIEDEL2013}, quantized nonlinear optical properties \cite{dejuanQuantizedCircularPhotogalvanic2017, orensteinTopologySymmetryQuantum2021a, maTopologyGeometryNonlinear2021}, and chiral anomalies \cite{zyuzinTopologicalResponseWeyl2012,zyuzinTopologicalResponseWeyl2012, barnesElectromagneticSignaturesChiral2016}, etc.

\begin{figure} 
    \includegraphics[width=8.6cm]{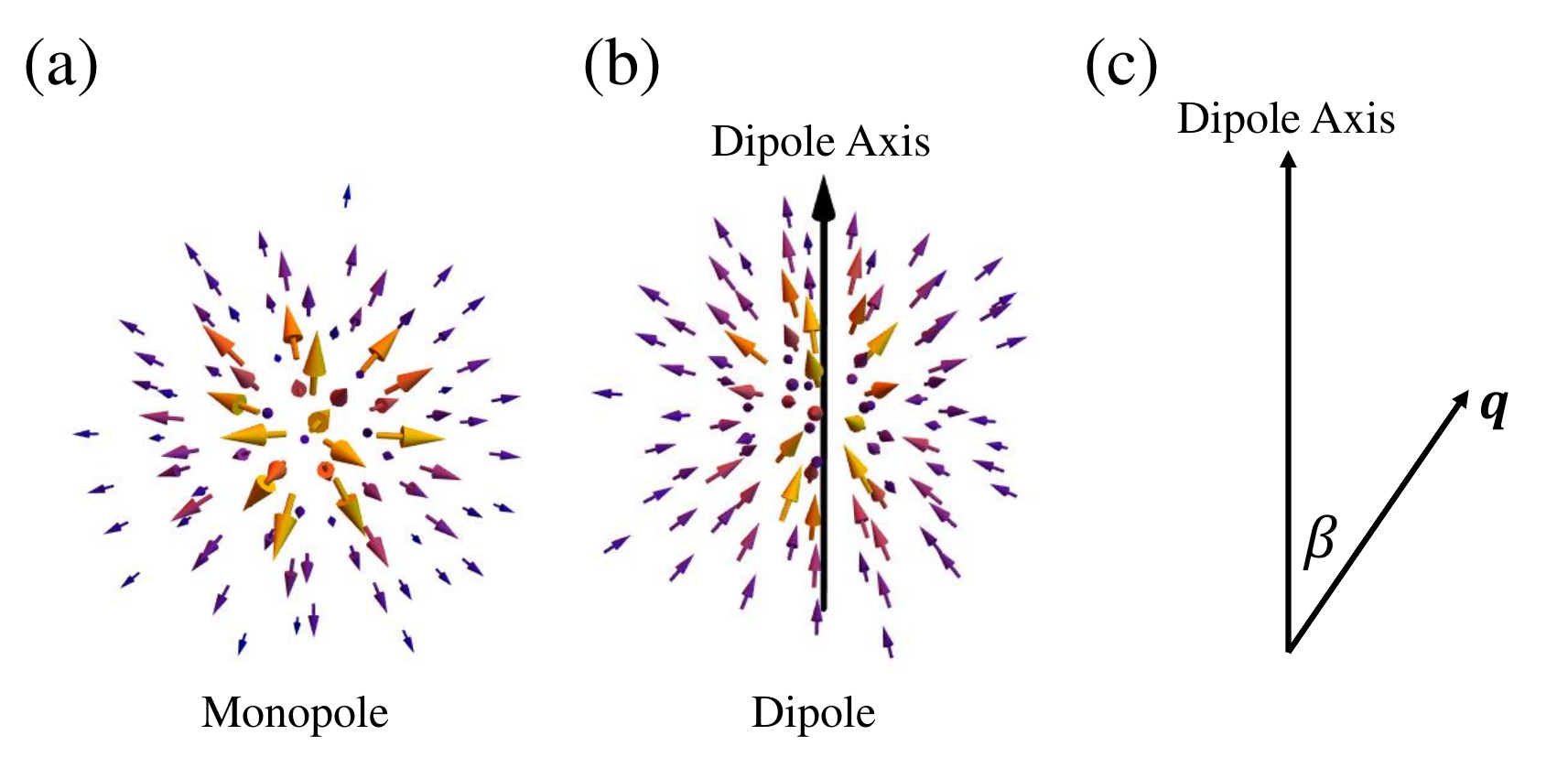}
        \caption{Schematic images of (a) monopole and (b) dipolar Berry curvature. (c) the plasmon wavevector $\bm q$ at the angle $\beta$ from the dipole axis.}
    \label{fig:Berry_cuvature}
\end{figure}

Recent studies have revealed new types of topological semimetals such as multi-Weyl semimetals \cite{fangMultiWeylTopologicalSemimetals2012}, nodal line semimetals \cite{fangTopologicalNodalLine2015, fangTopologicalNodalLine2016}, and chiral multifold semimetals \cite{bradlynDiracWeylFermions2016, changUnconventionalChiralFermions2017}. Similar to Dirac semimetals, these materials exhibit intriguing physical features driven by their band topological properties.
Recently, a novel class of topological semimetals, referred to as multifold Hopf semimetals, has been proposed \cite{grafMasslessMultifoldHopf2023}. Multifold Hopf semimetals host band crossings involving multiple isotropic linear and flat energy bands. For instance, the energy-band structure of threefold Hopf semimetals consists of a Dirac cone and a flat band, which is the same as that of chiral multifold semimetals. A distinguishing feature of multifold Hopf semimetals that separates them from chiral multifold semiemtals is that the band-crossing node acts as a point-like Berry dipole instead of a Berry monopole ( Fig.~\ref{fig:Berry_cuvature}), which can be mathematically expressed as
\begin{equation}
    \bm \Omega(\bm q)\propto (\bm d \cdot \bm  q)\frac{\bm  q}{\left| \bm  q \right|^4}.
\end{equation}
where $\bm d$ denotes the dipole vector \cite{grafMasslessMultifoldHopf2023}. It should be noted that, unlike the monopole Berry curvature, the dipolar Berry curvature is anisotropic and can thus induce anisotropy in the material properties despite the isotropic structure of the energy band. For example, a recent theoretical work shows that threefold Hopf semimetals exhibit anisotropic optical responses owing to different selection rules depending on the polarization axis of light \cite{habeOpticalConductivityThreefold2022, ahnMagnetoopticalConductivityThreefold2024}

Plasmons are self-sustaining, collective electron modes. Their classical behavior can be evaluated from a hydrodynamic perspective by solving the Euler equation of motion for the current density and continuity equation \cite{giulianiQuantumTheoryElectron2005}. It is worth noting that in the classical hydrodynamic approach, only energy-band dispersion is involved, which implies that the isotropy of classical plasmons is directly determined based on the isotropy of the energy band. Thus, according to classical intuition, plasmons of multifold Hopf semimetals should be isotropic due to their isotropic energy bands. However, as noted above, multifold Hopf semimetals possess internal anisotropy in their electronic wave functions. Therefore, caution should be exercised when concluding that plasmons are isotropically based solely on the energy band shape.

In this study, our focus is on threefold Hopf semimetals, where we investigate the effects of the anisotropic electronic wavefunction structure associated with the dipolar Berry curvature on plasmon anisotropy. We find that plasmons are anisotropic even in the classical long-wavelength regime despite the isotropy of the energy band, and the degree of anisotropy may reach up to 25\%, which is sufficiently large to be experimentally verified.

The paper is organized as follows. In Sec.~\ref{sec:model}, we review a minimal model for threefold Hopf semimetals. In Sec.~\ref{sec:polarizability}, we provide our formalism for the dynamic polarizability along with its numerical results. Section~\ref{sec:anisotropic plasmons} presents a detailed analytical and numerical analysis on plasmon anisotropy. We conclude in Sec~\ref{sec:conclusion}.

\section{Model} \label{sec:model}

The essential features of the low-energy excitations of threefold Hopf semimetals can be captured by the minimal Hamiltonian expanded around the nodal point, written as \cite{habeOpticalConductivityThreefold2022, grafMasslessMultifoldHopf2023}
\begin{equation} \label{eq:Hamiltonian}
H_{\bm q}= v
\left(\begin{array}{ccc}
0 & q_{-} & -i q_z \\
q_{+} & 0 & 0 \\
i q_z & 0 & 0
\end{array}\right)
\end{equation}
where $v$ is the Fermi velocity, $q_{\pm}=q_x \pm i q_y$, and $q = \left| \bm q\right |$ with $\bm q = (q_x, q_y, q_z)$. Here, we set the Berry dipole axis along the $z$-axis without a loss of generality.
The energy band dispersion of this minimal model is $\varepsilon_{s, \bm q}=s v q$, where $s=1,0,-1$ denotes the band index. It should be noted that the energy bands are identical to those of chiral multifold semimetals consisting of a Dirac cone and a single flat band. The electronic wavefunctions, however, possess a unique structure that is characterized by an anisotropic Berry curvature. The formula for Berry curvature is:

\begin{align} \label{eq:Berry_curvature}
    \Omega_{s,k}(\boldsymbol{q})=-\epsilon_{ijk}\sum_{s'\neq s}\frac{2\mathrm{Im}[\langle\psi_{s,\boldsymbol{q}}|v_i|\psi_{s',\boldsymbol{q}}\rangle\langle\psi_{s',\boldsymbol{q}}|v_j|\psi_{s,\boldsymbol{q}}\rangle]}{|{\varepsilon}_{s,\boldsymbol{q}}-{\varepsilon}_{s',\boldsymbol{q}}|^2},
\end{align}
where
\begin{equation}
    \begin{gathered}
        |\psi_{\pm,\boldsymbol{q}}\rangle=    
        \frac{1}{\sqrt{2} q}\left(\begin{array}{c}
        q \\ 
        \pm q_{+} \\
        \pm i q_z
        \end{array}\right) \\
        |\psi_{0,\boldsymbol{q}}\rangle=    
        \frac{1}{q}\left(\begin{array}{c}
        0 \\ 
        i q_z \\
        q_{-}
        \end{array}\right)
    \end{gathered}
\end{equation}
are the wavefunctions of the threefold Hopf semimetals obtained by diagonalizing Eq.~(\ref{eq:Hamiltonian}). By evaluating the Berry curvature using Eq.(\ref{eq:Berry_curvature}), we obtain 
\begin{equation} \label{eq:dipolar_Berry_curvature}
    \bm \Omega_s(\bm q)=\kappa_s (\hat{\bm d}\cdot \bm  q)\frac{\bm  q}{\left| \bm  q \right|^4},
\end{equation}
where $\hat{\bm d}=(0,0,1)$ denotes the unit vector in the direction of the dipole axis [ Figs. ~\ref{fig:Berry_cuvature}(b)], and $\kappa_s$ is the dipole charge associated with each energy band with $\kappa_{\pm}=-1$ and $\kappa_0=2$. It should be noted that Eq.~(\ref{eq:dipolar_Berry_curvature}) describes a singular point-like Berry dipole structure that is located at the band-crossing point, in contrast to the monopole structure of the Berry curvature that appears in Dirac materials.

\section{Dynamic polarizabiltiy } \label{sec:polarizability}
We first evaluate the zero-temperature dynamic polarizability of threefold Hopf semimetals by explicitly calculating 
\begin{equation} \label{eq:polarizability}
    \Pi_{ss'}(\bm q,\omega) = -\int \frac{d \bm k}{(2\pi)^3} \sum_{s, s'} \frac{n_\mathrm{F}(\xi_{s,\bm k})-n_\mathrm{F}(\xi_{s',\bm k+\bm q} )}{\omega+\varepsilon_{s, \bm k}-\varepsilon_{s', \bm k+\bm q}+i\eta} F^{ss'}_{\bm k, \bm k + \bm q}.
\end{equation}
Here, $\xi_{s,\bm k}=\varepsilon_{s,\bm k} -E_\mathrm{F}$ is the energy measured from the Fermi energy $E_\mathrm{F}$, $n_\mathrm{F}(-\xi_{s,\bm k})=\Theta(\xi_{s,\bm k})$ is the Fermi distribution function at zero temperature, where $\Theta$ is the Heaviside step function, $\eta$ is a phenomenological scattering rate with an infinitesimal value (i.e., $\eta \rightarrow 0^+$) in the clean limit, and $F^{ss'}_{\bm k, \bm k + \bm q}=\left| \braket{\psi_{s,\bm k}}{\psi_{s',\bm k + \bm q}} \right|^2$ is the square of overlap between $\ket{\psi_{s,\bm k}}$ and $\ket{\psi_{s',\bm k + \bm q}}$. The total polarizability is given by the summation of the band indices:
\begin{align} \label{eq:total_polarizability}
    \Pi(\bm q,\omega)&=\sum_{s, s'}\Pi_{ss'}(\bm q,\omega) \nonumber \\ 
    &=\sum_{s \geq s'} \Pi_{ss'}(\bm q,\omega)\delta_{ss'} + \overline{\Pi}_{ss'}(\bm q,\omega)(1-\delta_{ss'})
\end{align}
where $\overline{\Pi}_{ss'} = \Pi_{ss'}(\bm q,\omega) + \Pi_{s's}(\bm q,\omega)$.
The total polarizability can be decomposed into intraband and interband parts (expressed as $\Pi = \Pi_\mathrm{intra} + \Pi_\mathrm{inter}$), which are contributed only by intraband and interband transitions, respectively. It is clear that $\Pi_\mathrm{intra}(\bm q,\omega)=\Pi_{++}(\bm q,\omega)$ and $\Pi_\mathrm{inter}(\bm q,\omega)=\overline{\Pi}_{+-}(\bm q,\omega) + \overline{\Pi}_{+0}(\bm q,\omega)$. It should be noted that we neglect terms that involve transitions between occupied bands, such as $\Pi_{--}$, which should not be nonzero owing to Pauli blocking.

\begin{figure} 
    \includegraphics[width=8.6cm]{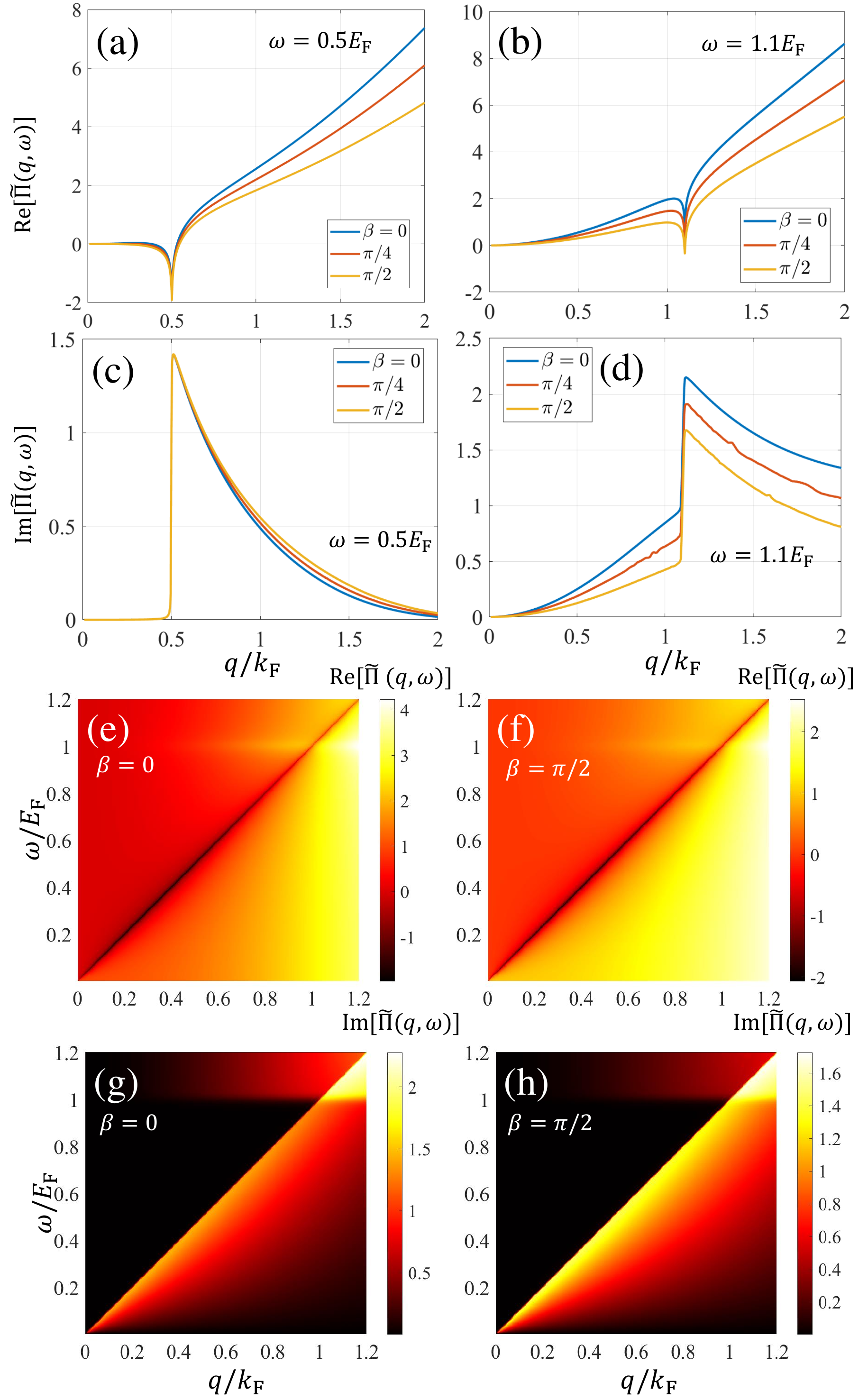}
        \caption{
        (a)-(d) Plots of dynamic polarizability as a function of the momentum magnitude $q$ along different momentum directions(i.e., for different values of $\beta$), showing the anisotropy of the polarizability. (e)-(f) Density plots of real and imaginary parts polarizability as a function of $\omega$ and $q$. Here, $\beta$ is the angle of the momentum $\bm q$ from the axial dipole axis [see Fig.~\ref{fig:Berry_cuvature}(c)] and the polarizability is normalized by the density of state at the Fermi energy, i.e., 
        $\widetilde{\Pi}=\Pi/D(\varepsilon_\mathrm{F})$ where $D(\varepsilon)$ is the density of states at the Fermi energy.
        }
    \label{fig:polarizability}
\end{figure}

Figure~\ref{fig:polarizability} presents the numerically calculated polarizabilities along different directions $\beta=0$, $\pi/4$, and $\pi/2$, where $\beta$ is the angle between the momentum $\bm q$ and the dipole axis of the Berry curvature [see Fig.~\ref{fig:Berry_cuvature}(c)].
For the numerical calculations, we set $\eta=10^{-3}E_\mathrm{F}$ to avoid difficulties arising from singularities during the integration of Eq.~(\ref{eq:polarizability}). While this small value of $\eta$ does blur sharp features of the polarizability, it remains negligible for our purposes, as it does not lead to qualitatively or quantitatively significant deviations from the results obtained in the clean limit.
Figs.~\ref{fig:polarizability}(a)–(d) present the real and imaginary parts of the polarizabilty as functions of $q$ at a fixed $\omega$. For $\omega=0.5 E_\mathrm{F}$[Figs.~\ref{fig:polarizability}(a) and (c)], the real and imaginary parts of the polarizabilty are close to zero for small values of $q<0.5k_F$, and thus, they exhibit little anisotropy. At $q=0.5k_F$, the real(imaginary) part exhibits a sharp dip(abrupt increase) owing to the onset of interband transitions from the Dirac cone valence band($s=-1$) to the Dirac cone conduction($s=+1$) band. In the region $q>0.5k_F$ where interband transitions are available, the polarizability exhibits a larger anisotropy as $q$ increases.
For $\omega=1.1 E_\mathrm{F}$[Figs. ~\ref{fig:polarizability}(b) and (d)], the real and imaginary parts of the polarizabilty are nonzero even for a very small $q$. This is because interband transitions from the flat valence band ($s=0$) to the Dirac cone conduction band ($s=1$) are available at any value of $q$ when $\omega>E_F$. Therefore, the polarizability is highly anisotropic over a wide range of $q$, with singular behaviors appearing at $q=1.1E_F$ owing to interband transitions from the Dirac cone conduction($s=-1$) to Dirac cone valence($s=+1$) bands.

Figures~\ref{fig:polarizability}(e)–(h) illustrates the density plot of the polarizability along various directions of the momentum wavevector $\bm q$.
The polarizability is similar to that of Weyl semimetals due to the linear energy band structure of the upper conduction and lower valence energy bands. It is worth noting that the single-particle excitation regime, which is defined as the ($q$, $\omega$) space wherein $\operatorname{Im}\Pi\neq0$, appears exactly the same at $\omega<E_\mathrm{F}$ \cite{lvDIELECTRICFUNCTIONFRIEDEL2013}. However, additional interband transitions from the occupied flat band to the unoccupied conduction band are available for threefold Hopf semimetals, which subsequently expands the single-particle excitation region (where $\operatorname{Im}\Pi\neq0$) to cover the area $\omega>E_F$ for all values of $q$ as shown in Figs. ~\ref{fig:polarizability}(g) and (h).

\begin{figure*} 
    \includegraphics[width=12cm]{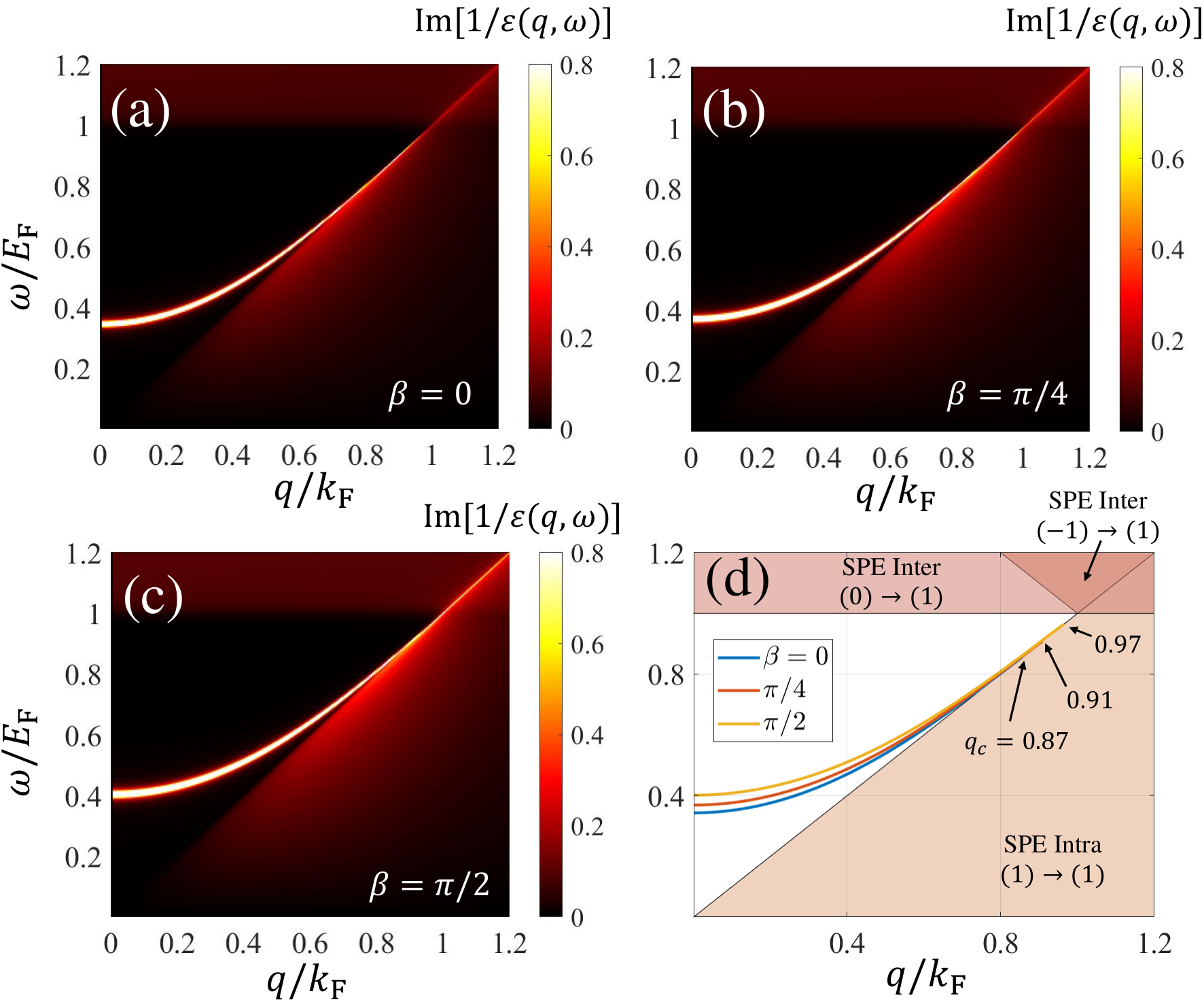}
        \caption{(a)-(c) Density plots of the loss function at different values of $\beta$, defined as the imaginary part of the inverse of the dielectric function. (d) The plasmon dispersions along different directions (i.e., at different values of $\beta$) with $q_\mathrm{c}$ indicating the critical momenta where the plasmon dispersion merges into the single-particle excitation (SPE) regime with plasmons decaying through Landau damping. }
    \label{fig:plasmons}
\end{figure*}
\section{anisotropic plasmons} \label{sec:anisotropic plasmons}
Plasmons are self-sustaining collective excitations that are accompanied by charge density oscillations. The plasmon energy dispersion is determined by finding the zeros of the dielectric functions \cite{giulianiQuantumTheoryElectron2005}. 
Within the random-phase approximation, the dielectric function can be obtained using the polarizability given by $\varepsilon(\bm q,\omega)=1+v_q\Pi(\mathbf{q}, \omega)$
where $v_q=\frac{4\pi e^2}{\kappa q^2}$ is the Coulomb interaction, and $\kappa$ is the background dielectric constant. Thus, we can obtain the plasmon dispersion $\omega_\mathrm{p}(\bm q)$ by solving the following equation:
\begin{equation} \label{eq:plasmons_zeros}
    \varepsilon[\bm q,\omega_\mathrm{p}(\bm q)]=0.
\end{equation}
Similar to Weyl semimetals or graphene, the strength of Coulomb interactions is characterized by an effective fine structure $\alpha=e^2/\kappa v$. Thus, the dielectric function can then be rewritten as
\begin{equation} \label{eq:dielectric_function}
\varepsilon(\bm q,\omega)=1+\frac{2\alpha}{\pi \widetilde{q}^2} \widetilde{\Pi}(\mathbf{q}, \omega),
\end{equation} 
where $\widetilde{\Pi}(\mathbf{q}, \omega)=\Pi(\mathbf{q}, \omega)/D(\varepsilon_\mathrm{F})$ with $D(\varepsilon_\mathrm{F})=\frac{k_\mathrm{F}^3} {2\pi^2\varepsilon_\mathrm{F}}$ being the density of states and $\widetilde{q}=q/k_\mathrm{F}$.

Figures~\ref{fig:plasmons}(a)–(c) illustrate the frequency-dependent loss function, which is defined as the imaginary part of the inverse of the dielectric function (i.e., $\mathrm{Im}[\varepsilon^{-1}(\bm q, \omega)]$). The loss functions are calculated using Eq.~(\ref{eq:dielectric_function}) with $\alpha=2.0$. The results for the dynamic polarizability are presented in Fig.~\ref{fig:polarizability}. 
Plasmons of threefold Hopf semimetals exhibit typical three-dimensional plasmon behavior: the plasmon energy is finite at $q=0$ and increases with $q$. At the critical momentum $q_c$, the plasmon energy dispersion enters the single-particle excitation region [see Fig.~\ref{fig:plasmons}(d)], and thus, they rapidly decay through the generation of electron–hole pairs. It should be noted that while the plasmon linewidth is almost the same for all values of $\theta$, the plasmon energies are redshifted as $\theta$ decreased, as shown in Fig~\ref{fig:plasmons}(d). Interestingly, the plasmon dispersion exhibits stronger anisotropy at a smaller $q$, in contrast to our intuition that the anisotropy disappears in the limit $q\rightarrow 0$, i.e., classical regime.

In the following, we investigate the origin of the anisotropy of polarizability by obtaining the analytical form of the plasmon frequency[i.e., $\omega_\mathrm{p}(q\rightarrow 0)$] in the long-wavelength limit. We begin by expanding Eq.~(\ref{eq:polarizability}) in powers of $q$ around $q=0$: The overlap factor $F^{ss'}_{\bm k, \bm k + \bm q}$ is approximated up to the second order of $q$ as follows:
\begin{widetext} 
    \begin{equation} \label{eq:overlap_exp}
        F^{ss'}_{\bm k, \bm k+ \bm q} \approx
        \begin{cases}
            1-\frac{q^2}{4 k^2}\left(A_{ss'}+S_{ss'} \sin 2 \beta+C_{ss'} \cos 2 \beta\right), & s=s' \\ 
            \frac{q^2}{4 k^2}\left(A_{ss'}+S_{ss'} \sin 2 \beta+C_{ss'} \cos 2 \beta\right), & s \neq s'
        \end{cases}
    \end{equation}
\end{widetext}
where
$A_{++}=1+A_{-+}$, $S_{++}=-\sin{2\theta}\cos{\phi}$, $C_{++}=-\frac{1}{2}C_{00} - A_{-+}$,
$A_{00}=2$,  $S_{00}=2S_{1,1}$, $C_{00}=-2\cos{2\theta}$,
$A_{-+}=\frac{1}{2}\sin^2{\theta}\sin^2{\phi}$, $C_{-+}=-A_{-+}$, $S_{-+}=0$,
$A_{0+}=1$, $C_{0+}=-C_{++}$ and $S_{0+}=S_{++}$. Here, we use the spherical coordinates with $k_z$($k_x$) direction corresponding to $\theta=0$($\phi=0$). The remaining components can be easily derived from the observation that $F^{ss'}_{\bm k, \bm k+q}$ remains unchanged when swapping the indices (i.e., $s\leftrightarrow s'$), and $F^{1,1}_{\bm k, \bm k+q}$ is equivalent to $F^{-1,-1}_{\bm k, \bm k+q}$ because of the electron-hole symmetry of the system. 
By inserting Eq.(\ref{eq:overlap_exp}) into Eq.~(\ref{eq:polarizability}) and expanding it to the second order in terms of $q$, we obtain the following analytical form of the long-wavelength polarizability
\begin{align} 
    \widetilde{\Pi}_{++}(q, \omega; \beta)&=
        -\frac{\widetilde{q}^2}{3 \widetilde{\omega} ^2} \label{eq:polarizability_approx1} \\ 
    \widetilde{\overline{\Pi}}_{+-}(q, \omega; \beta)&=
        -\frac{1}{24} \widetilde{q}^2 \log \left(\frac{\widetilde{\omega}^2-4}{\widetilde{\omega} ^2-4 \widetilde{\Lambda}^2}\right) \sin^2{\beta} \label{eq:polarizability_approx2} \\
    \widetilde{\overline{\Pi}}_{+0}(q, \omega; \beta)&=
        -\frac{1}{6} \widetilde{q}^2 \log \left(\frac{\widetilde{\omega} ^2-1}{\widetilde{\omega} ^2-\widetilde{\Lambda}^2}\right) \left(\cos^2 {\beta} +1 \right) \label{eq:polarizability_approx3}
\end{align}
where $\widetilde{\omega}=\omega/E_\mathrm{F}$, $\widetilde{\Lambda}=\Lambda/k_\mathrm{F}$, and $\Lambda$ is the cutoff momentum of the inverse lattice spacing order. 
From the above equations, it can be observed that the interband part of the polarizability $\Pi_\mathrm{intra}(\bm q,\omega)=\Pi_{++}(\bm q,\omega)$ is totally isotropic without dependence on the direction of $\bm q$, while $\Pi_\mathrm{inter}(\bm q,\omega)=\overline{\Pi}_{+-}(\bm q,\omega) + \overline{\Pi}_{+0}(\bm q,\omega)$ is anisotropic with explicit dependence on $\beta$. It is worth noting that $\overline{\Pi}_{+-}$ is zero for $\beta=0$, that is, along the direction of the dipole axis. To gain a deeper understanding of this, we approximate the overlap factor $\ket{\psi_{s,\bm k + \bm q}}$ as $\partial_{\bm k}\ket{\psi_{s,\bm k}}\cdot \bm q$ in the long-wavelength limit($q \ll 1$) and rewrite the overlap factor as
\begin{align} \label{eq:overlap_approx}
    F^{+-}_{\bm k, \bm {k+q}} &\approx\left|\matrixel{\psi_{+,\bm k}}{\bm q \cdot \partial_{\bm k}}{\psi_{-,\bm k}}\right|^2 \nonumber \\
    &= \sum_i  q_i^2 \left|\matrixel{\psi_{+,\bm k}}{ \partial_{k_i}}{\psi_{-,\bm k}}\right|^2  \nonumber \\
    &= \sum_i  q_i^2 \left|\frac{\matrixel{\psi_{+, \bm k}}{v_i}{\psi_{-, \bm k}}}{\varepsilon_{-,\bm k}-\varepsilon_{+,\bm k}}\right|^2
\end{align}
where $i=x,y,z$ is the coordinate, and $v_i=\partial H/\partial k_i$ is the velocity operator. 
In the presence of particle–hole and mirror symmetries, which protect the band-crossing point node in our minimal model[Eq.(\ref{eq:Hamiltonian})],
$\matrixel{\psi_{-,k}}{v_i}{\psi_{+,k}}=\xi_i \matrixel{\psi_{-,k}}{v_i}{\psi_{+,k}}$, where $\xi_i=1$ for $i=x,y$ and $\xi_i=-1$ for $i=z$ \cite{habeOpticalConductivityThreefold2022}. This immediately leads to $\matrixel{\psi_{-,k}}{v_z}{\psi_{+,k}}=0$, indicating that the overlap factor should also be zero (and thus intercone transitions should be forbidden) when the momentum wavevector $\bm q$ aligns with the dipole axis [that is, $\bm q = (0, 0, q_z)$].

The dielectric function of the threefold Hopf semimetals in the long-wavelength limit can be obtained by substituting Eqs.~(\ref{eq:polarizability_approx1})–(\ref{eq:polarizability_approx3}) into Eq.~(\ref{eq:dielectric_function}). In the presence of only intraband transitions, we can analytically obtain the dielectic function by substituting $\Pi(\bm q, \omega)$ in Eq.~(\ref{eq:dielectric_function}) by $\Pi_\mathrm{intra}(\bm q, \omega)$. Using Eq.~(\ref{eq:plasmons_zeros}), the long-wavelength plasmon frequency is given by $\omega_\mathrm{p}=\sqrt{\frac{2\alpha}{3\pi}}$ and has no directional dependence. The full plasmon frequency, which includes interband transitions, is not analytically solvable and requires numerical computations. To obtain insights into the direction-dependent behavior of the plasmon present in Fig.~\ref{fig:plasmons}, we consider the limits of small $\alpha\ll 3\pi/(4\log{\widetilde{\Lambda}})$ and small $\omega\ll E_\mathrm{F}$. We can then obtain the analytical form of the long-wavelength plasmon frequency in the presence of interband transitions, given by
\begin{equation} \label{eq:plasmon_longwavelength}
    \omega_\mathrm{p}(\beta)=\omega_\mathrm{p}^\mathrm{intra}+\sum_{n=1}^\infty \frac{(2n-1)!!}{n!(-2)^n}A(\beta)^n\left(\frac{2\alpha}{3\pi} \right)^{n+\frac{1}{2}}
\end{equation}
where $A(\beta)=\frac{13+3\cos{2\beta}}{8}\log{\frac{\Lambda}{E_\mathrm{F}}}$, The formula for the plasmon frequency[Eq.~(\ref{eq:plasmon_longwavelength})] shows that the long-wavelength plasmon frequency decreases as $\beta$ increases, which is consistent with the previously discussed numerical results. It should be noted that the anisotropy is stronger for a large interaction strength($\alpha$) and a large cut-off momentum($\Lambda$).

\begin{figure} 
    \includegraphics[width=8.6cm]{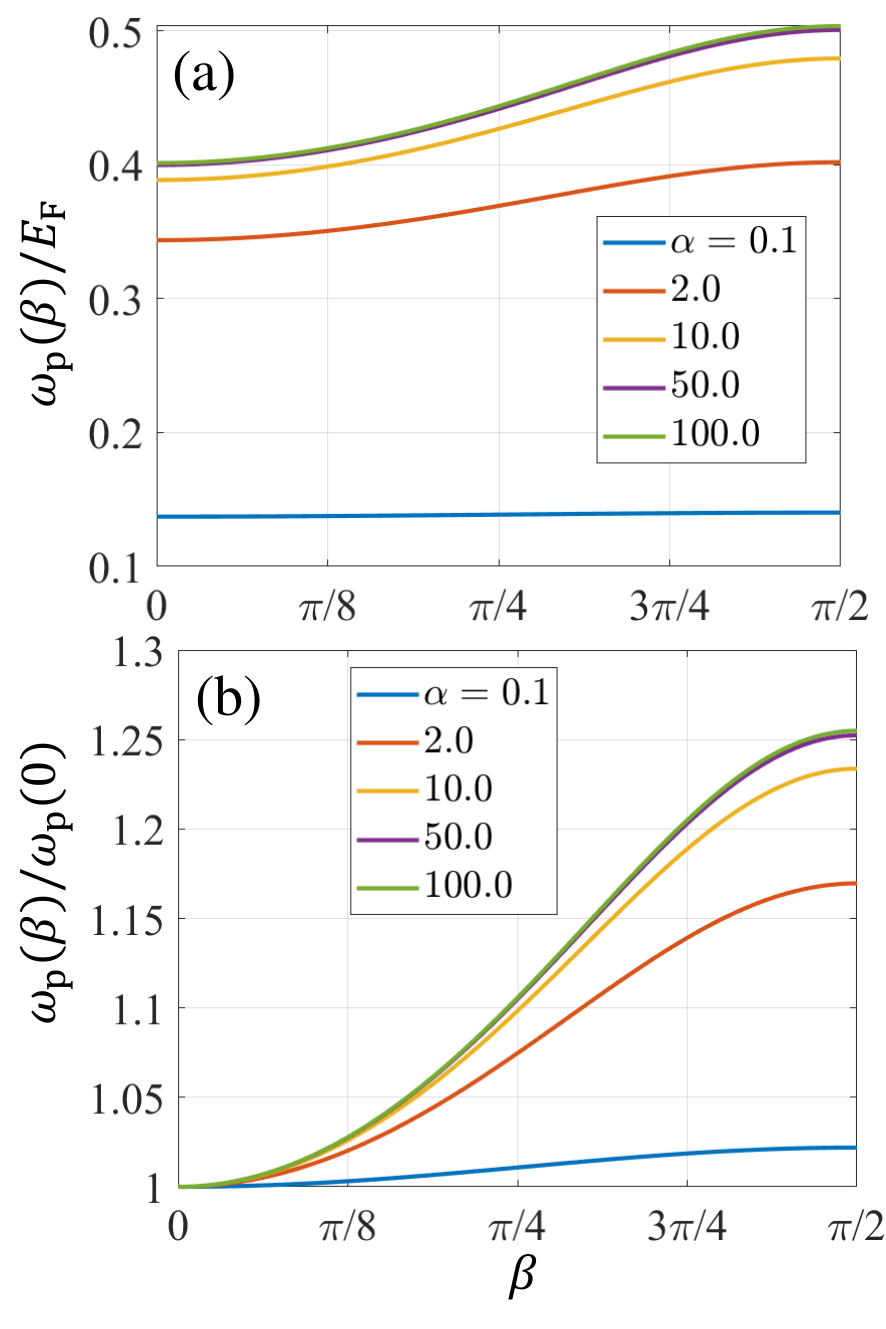}
        \caption{Long-wavelength plasmon frequency as a function of $\beta$ for different values of $\alpha$, which characterizes the strength of Coulomb interactions. Here, the plasmon frequency is normalized by (a) the Fermi energy and (b) the plasmon frequency at $\beta=0$. }
    \label{fig:plasmons_longwavelnegth}
\end{figure}

To verify the validity of the analysis described earlier, the long-wavelength plasmon frequency is numerically calculated as a function of $\theta$ at various interaction strengths in Fig.~\ref{fig:plasmons_longwavelnegth}. It should be noted that the long-wavelength plasmon frequency is almost isotropic for a very small value of $\alpha\ll 1$ with $\gamma=\omega_\mathrm{p}(\pi/2)/\omega_\mathrm{p}(0)\sim 1.02$. The anisotropy increases at larger values of $\alpha$ with $\gamma$ reaching approximately $1.25$ when $\alpha=100$. Although our theoretical predictions for such a large value of $\alpha$ may not be valid because the perturbative random phase approximation employed in the calculations breaks down when the interaction strength is large ($\alpha \gg 1$) \cite{giulianiQuantumTheoryElectron2005}. However, the results presented in Fig.~\ref{fig:plasmons_longwavelnegth} clearly show that for values of $\alpha$ on the order of unity, where the random phase approximation is valid, the anisotropy factor $\gamma$ is in the range of  $1.2 > \gamma > 1.1$, which is sufficient to be observed experimentally.

\begin{figure} 
    \includegraphics[width=8.6cm]{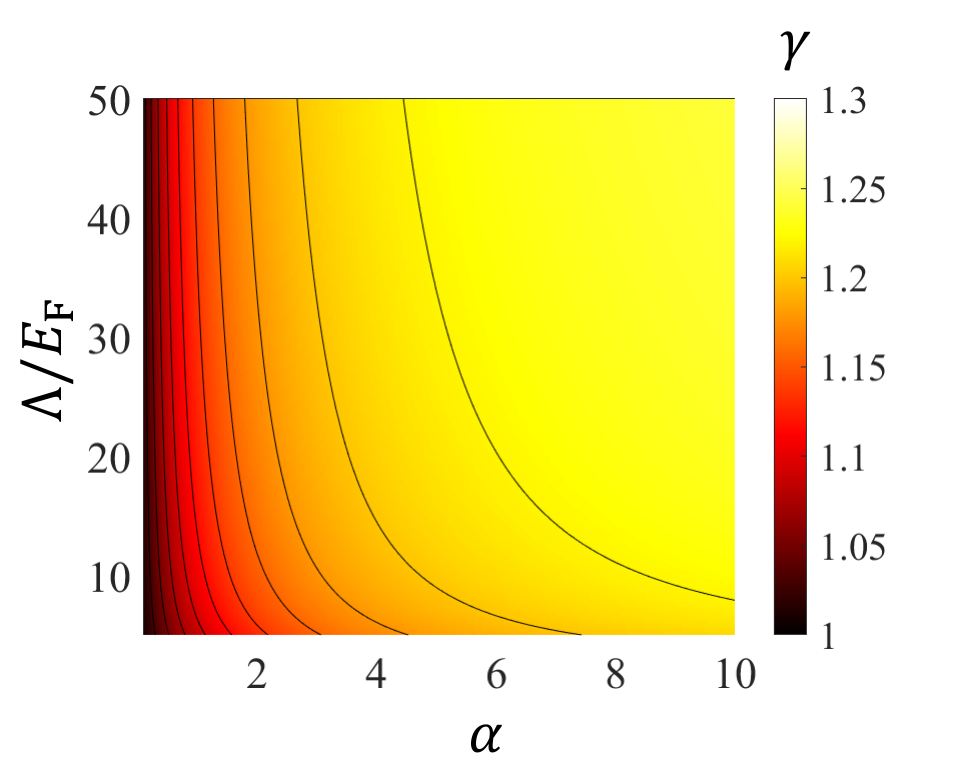}
        \caption{A density plot of the ratio $\gamma$ as a function of the cutoff momenta $\Lambda$ and the interaction strength $\alpha$. Here $\gamma=\omega_\mathrm{P}(\beta=\pi/2)/\omega_\mathrm{P}(0)$, which is the ratio between the plasmon frequencies at $\beta=0$ and $\beta=\pi/2$, characterizing the anisotropy of long wavelength plasmons. }
    \label{fig:plasmons_longwavelnegth_alpha_lambda}
\end{figure}

In Fig.~\ref{fig:plasmons_longwavelnegth_alpha_lambda}, we investigate the impact of the interaction strength and cutoff momentum on the anisotropy of the plasmon by providing a density plot with contour lines of the ratio $\gamma$ in a reasonable range of the interaction strength $\alpha$ and cutoff momentum $\Lambda$. 
It should be noted that while the plasmon frequency is significantly affected by the interaction strength, its dependence on the cutoff momentum is much weaker owing to the logarithmic nature of this dependence [see Eq.~(\ref{eq:plasmon_longwavelength})]. In addition, it should be noted that the anisotropy ratio $\gamma$ can be large, up to $1.25$ across the entire parameter space, indicating that the plasmon anisotropy of threefold Hopf semimetals should be sufficiently large to be experimentally verified.

\section{ Conclusion } \label{sec:conclusion}
In this study, we investigated the anisotropy of plasmons in threefold Hopf semimetals. We found through the integration of the random phase approximation that the internal anisotropy of threefold Hopf semimetals, which is associated with the dipolar Berry curvature, induces plamon anisotropy with a degree reaching up to 25\%. Our analysis demonstrates that plasmons exhibit a strong directional dependence, with anisotropy becoming more pronounced at higher interaction strengths and cutoff momenta. These findings are crucial for understanding plasmons in threefold Hopf semimetals and will be essential for future theoretical and experimental studies involving plasmons. Our work will also aid the experimental identification of threefold Hopf semimetals using experimental techniques such as electron energy-loss spectroscopy. 

\begin{acknowledgments}
This work was supported by the National Research Foundation of Korea(NRF) grant funded by the Korea Government (MSIT) (No. RS-2023-00272513).
\end{acknowledgments}

\end{document}